\documentclass[10pt]{article}
\usepackage{epsfig,graphics,color}
\author{\large \bf  Z.~Usubov\footnote
         {On leave of absence from Institute of Physics, Baku, Azerbaijan}
\\
\\Joint Institute for Nuclear Research,
\\ Dubna, Russia}          
\title { Dijet Signature of Low Mass  Strings in the \\      
                           Early LHC Data }              

\begin{document}
\maketitle
{
\vskip 0.5cm
\hskip 6.5cm {\bf \large { Abstract}}
\vskip 1.0cm

{\large {
We have examined the dijet production  at  the LHC 
as a hint to discovery of string Regge excitations of Standard
Model particles.
If the fundamental string mass scale is in the TeV range, 
the influence of string effects on the $pp$ interaction
can arise as new resonances in the dijet invariant mass distribution.
Near the first resonant pole we investigate  the impact of the subprocesses
with quark-gluon and four-gluon separately.
We show that the LHC                        
is able to probe   the string mass scale $\sim$5 TeV
with only 2 pb$^{-1}$ of integrated luminosity at $E_{CM}=14$ TeV.

\Large {
\section{Introduction}
$ $

\vskip -0.5cm
Daily operation of the LHC brings us closer to the discovery of new physics.

Many new physics scenarios -- supersymmetry, large or warped
extra dimensions, technicolor,
little Higgs models, etc.-- offered a solution 
of existing unsolved problems of the Standard Model(SM)       
around  the TeV mass scale. 

The space-time with more than four dimensions\cite{anto0} can explain the
%%%Existence of large extra dimensions\cite{anto0} can explain the
weakness of gravity and resolve the hierarchy problem\cite{arka1}, 
which is one of the main motivations for physics beyond the SM.
The goal is achieved by adding $N_{ED}$  flat large  
extra dimensions of space,
whichs are transparent only for gravitons.
The Planck mass $M_P$ and fundamental gravitational scale $M$ 
in this case  are related as
\begin{equation}
 M_{P}^{2}= {8 \pi} M^{2+N_{ED}}R^{N_{ED}},   
\end{equation}
where all $N_{ED}$ flat  extra dimensions are compactified to a
radius $R$. If one puts $M \sim \cal O$(1 TeV) to avoid the 
hierarchy problem, $R$ becomes very large for $N_{ED}=1\,(R \sim 10^8$ km)
and varies from $\sim$0.1 mm to a few fm when $N_{ED}$ ranges from
2 to 7. This speculation  leads to the distortion of the usual 
inverse square law of gravity at $r < R$. 
Within     this scenario it was shown that the tower
of graviton states with nonzero momentum (Kaluza-Klein(KK)
excitations) couples to SM particles\cite{giud,han,hewe,mira}.
The summation over  the 
whole tower of KK states makes graviton interactions with SM  
particles stronger. This leads to the observable effects in 
SM processes.                                              

Global fits to electroweak observables provide         
 lower bounds on $1/R$ in the 2-5 TeV range\cite{delga}.
Experimentally allowed values 
for the compactified radius of extra dimensions must be smaller 
than 44 $\mu$m\cite{kea}, which implies that the fundamental      
gravitational scale is larger than 3.2 TeV.

Since the mid-1980s the string theory (see, e.g. \cite{pol1,blum}      
and references therein) has been  regarded  as an 
elegant theory of Nature which is able to unify 
gravity and all other fundamental interactions\footnote
{Some string theory criticism  can be found in\cite{emam}
and references therein.}.
In the string theory everything is made up  of vibrating
strings and D-branes are  dynamical hypersurfaces that play an
essential role in    building models of particle interactions.
The string tension   ${\alpha}'$ 
satisfies  the linearly rising Regge trajectory
\begin{equation}
  j(s) = j_{0} + {\alpha}' s,    
\end{equation}
for recurrences with the angular momentum $j$ and 
the square of  the center-of-mass(CM)  energy~$s$.
The string theory is 
naturally based on the extra dimensions of space.

The scenario with large enough transverse extra dimensions\footnote
{(transverse to the branes, the string has Dirichlet  
boundary conditions)}
(see, e. g.  \cite{anto1}) implies  that  the 
fundamental string mass scale 
\begin{equation}
 M_S^{2}={\alpha}'^{-1}
\end{equation}
is  also of the order of few TeVs{\footnote   
{
It was shown that low scale nonsupersymmetric string models 
can radiatively \\  $\mbox{\quad\,\,}$  provide the generation  of electroweak
symmetry breaking\cite{anto2}.}}. 

If the string mass scale  is close to 1 TeV,
the tower of string excitation states of the  
graviton and SM particles
will open at the energies accessible to the LHC.       
To date,  there is no experimental evidence that the string 
theory itself  has  relation to the  
correct description of Nature.
The detection of these states 
would open the door to the testability of the string theory.

Direct searches for decays of string resonances 
into quark-quark, quark-gluon and  gluon-gluon 
pairs at the CMS\cite{cms1} at the integrated 
luminosity of 2.9~pb$^{-1}$
have ruled out the masses $M_S<2.5\,$TeV at the 95$\%$        
confidence level.                

The rest of this paper  is organized as follows. The next section 
gives a brief description of string phenomenology in parton-parton
interactions. Section~3 gives the description of signal
and background events modeling in the generic LHC detector. 
In Section~4 we present our results on the dijet signature of
string Regge excitations. We also make an estimate of the 
signal-to-background ratio for the  early stage of LHC running.
We end with the conclusions
in Section~5.

\section{String point of view on parton scattering processes}
$ $

\vskip -0.5cm
In this analysis we explore the dijet production at LHC
energies affected by the TeV scale string.
The processes we consider are
\begin{equation}
qq \to qq, 
\end{equation}
\begin{equation}
q \bar q \to q \bar q,
\end{equation}
\begin{equation}
gg \to q \bar q, 
\end{equation}
\begin{equation}
q \bar q \to gg,     
\end{equation}
\begin{equation}
qg \to qg,
\end{equation}
\begin{equation}
gg \to gg.
\end{equation}

Comprehensive investigation of string parton-parton scattering 
amplitudes was done in\cite{lust1,lust2}. 
The main concept of the model considered is the extensions of the
SM based on open strings ending on D-branes.              
The interactions occur between gauge bosons as strings
attached to stacks of D-branes and chiral matter as    
strings stretching between intersecting D-branes\cite{blum}.
All relevant tree level 2$\to$2 scattering amplitudes 
of the SM particles were  computed
with the requirement that the string theory is weakly coupled
and on the assumption of the low string mass scale.
It was shown that the amplitudes of the                            
processes involving four 
gauge bosons or two bosons and two fermions     
had the  compactification-model-independent, universal character.
These amplitudes only depend on the local intersection
properties of D-brane stacks and 
string effects  lead  to the                
common Veneziano formfactors               
\begin{equation}
 V(\hat s,\hat t,\hat u,M_{S}) \sim 
{ {\Gamma (1-\hat s/M_{S}^2) \Gamma(1-\hat u/M_{S}^2)}
\over {\Gamma (1 + \hat t/M_{S}^2)}} 
\end{equation}
which correspond  to an  infinite sum over  $s$-channel                   
poles at the masses of the string Regge excitations.
Here $\hat s,\,\hat t,\,\hat u$ are the Mandelstam 
invariants for the subprocesses.
It was pointed out\cite{lust1}  that at the partonic CM
energies $\sqrt{\hat s} \ll M_S$  the formfactors
$ V(\hat s,\hat t,\hat u,M_{S}) \sim 1-{{\pi}^2 \over 6} 
{\hat s} {\hat u}/M_S^4$
and the string contributions are suppressed.

At the partonic CM  energies above the low  string mass      
scale, $\sqrt{\hat s} > M_S$, the lowest    
Regge recurrences of SM particles can be manifested by contributions
to the 2$\to$2 SM parton cross sections. But even at the partonic
CM energies below the threshold virtual Regge                      
recurrences will impact on the SM  parton interaction cross sections.          

The exchange of graviton KK excitations in dijet production 
at $pp$ interactions  occurs  at the next order  
in perturbation theory and hence is significantly suppressed with 
respect to the string exchange\cite{peskp}.

\section{Signal and  background simulations for low string
mass scale in dijet events}
$ $

\vskip -0.5cm
In order to study the ability of the LHC 
to observe string dijet events in $pp$ collisions 
at  $E_{CM}=7$ and 14 TeV  we
used the PYTHIA6.4\cite{pyt1} event generator.
As an SM background, the partonic subprocesses   
(4)-(9)  were enabled.
The signal events also involve  subprocesses (4)-(9),
where the string influence on               
(6)-(9) was incorporated in PYTHIA over                                           
squared amplitudes at the leading order in string perturbation
theory\cite{lust1,lust2}. Thus, the SM dijet events involving
the contributions from string Regge 
recurrences will be regarded   as signal events.
Note that the ratio of the cross section of 
subprocesses (4)-(5) to the
cross section of (4)-(9) for the leading-order SM predictions
is 0.091(0.061) for $E_{CM}=7(14)$ TeV.
The initial-  and final-state 
QCD and QED radiation  
and multiple interactions   were enabled.
We used the leading-order 
parton distribution function set 
from  CTEQ6L1\cite{pump}.           

The LHC detector performance was simulated by using the 
publicly available PGS-4\cite{conw} 
with the parameter chosen to mimic a generic ATLAS-type
detector.
Jets were reconstructed down to $|\eta|\le 3$ using 
the anti-$k_T$ algorithm\cite{salam} which we  implemented in PGS-4.
We chose $D=0.7$ for the jet resolution parameter and required that both
leading jets carried a transverse momentum larger than
$p_{T,min}^{1,2}>100\,$GeV.
This puts  the lower limit  
$\sqrt{\hat s_{min}}=200\,$GeV.

We use the simplified
output from PGS-4, namely,  a list of two most energetic
jets.

We simulated each signal event set at the rates of 
 $\sim$2.2$\cdot$$10^7$.  5.2(3.2)$\cdot$$10^7$ SM
background events were simulated for $E_{CM}=7(14)$ TeV.
The data were  normalized to       
100 and 20 pb$^{-1}$ 
for  $E_{CM}=7$  and 14 TeV, respectively.
%%%%%%%%%%%%%%%%%%%%%%%%%%%5
%%\begin{figure}[htbp]
\begin{figure}[h!]                             %%   [H]      
\begin{center}
{\vskip -0.0cm}
{\hskip  0.0cm}  {\epsfxsize  5.0 truein \epsfbox{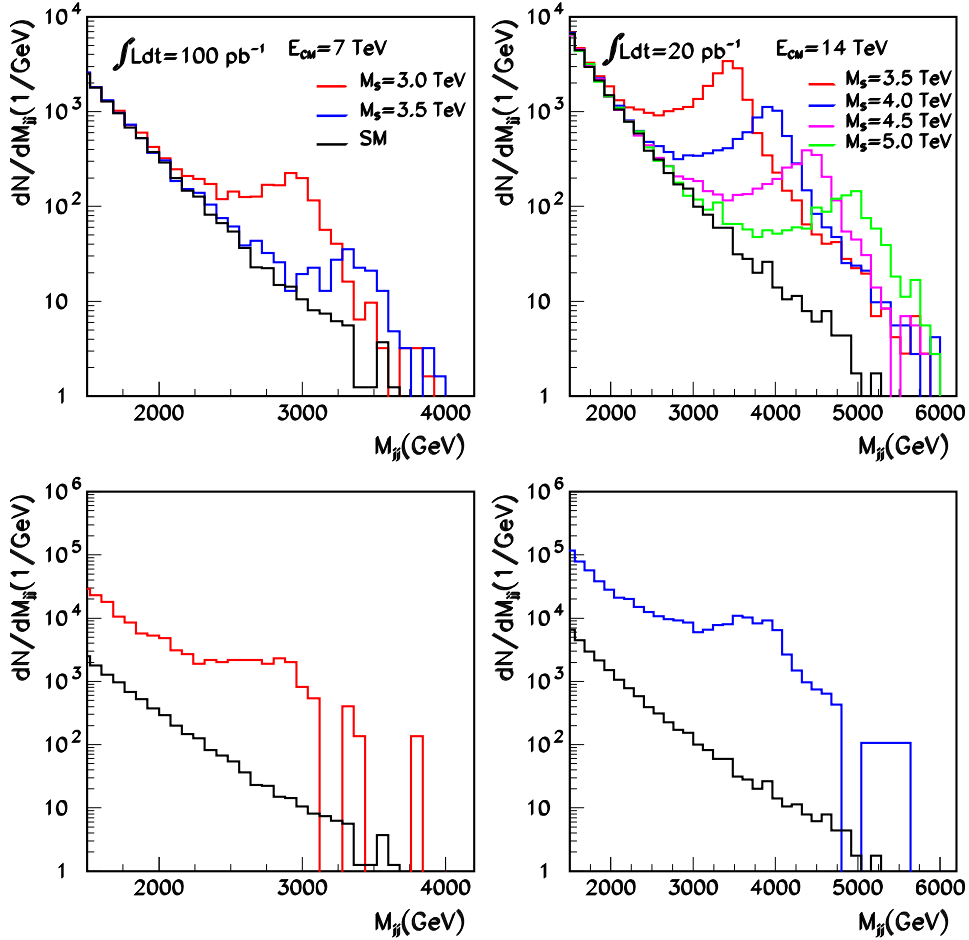}}
%%{\vskip -5.0cm}
\caption[]{\large {The dijet invariant mass distributions in $pp$ interactions
with quark-gluon(upper panels) and four-gluon(lower panels) subprocesses for
different values of $M_S$.
$\hat s \ge {\hat s}_{min}$ (see the text). The left(right) panels
corresponds to $E_{CM}=7(14)$ TeV. The
Standard Model prediction is shown in all panels(the black histogram).}}
\label{Norm}
\end{center}
\end{figure}

%%%%%%%%%%%%%%%%%%%%%%%%%%%5
%%\begin{figure}[htbp]
\begin{figure}[h!]                             %%   [H]      
\begin{center}
{\vskip -2.0cm}
{\hskip  0.0cm}  {\epsfxsize  5.0 truein \epsfbox{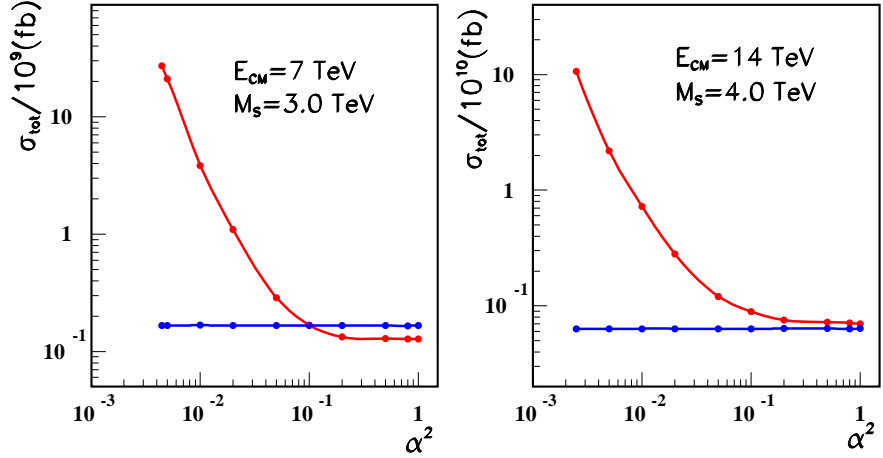}}
{\vskip -5.0cm}
\caption[]{\large {The expected  total cross section of $pp$ interactions 
with quark-gluon(blue) and four-gluon(red) subprocesses
as a function of ${\alpha}^2$(see the text).
The left(right) panel corresponds to $E_{CM}=7(14)$~TeV and $M_S=3(4)$~TeV.}}
\label{Norm}
\end{center}
\end{figure}

\section {Results}
$ $

\vskip -0.5cm
We first perform a comparison of the dijet invariant mass
$M_{jj}$$=$$\sqrt{(E_1+E_2)^2-(\vec p_1 + \vec p_2)^2}$
distributions of the signal and  the  SM background.
The system was composed of two leading  jets in $pp$ collisions
at the LHC with subprocesses (4)-(9).                 
%%%%%%%%%%%%%%%%%%%%%%%%%%%5
%%\begin{figure}[htbp]
\begin{figure}[h!]                             %%   [H]      
\begin{center}
{\vskip -2.0cm}
{\hskip  0.0cm}  {\epsfxsize  5.0 truein \epsfbox{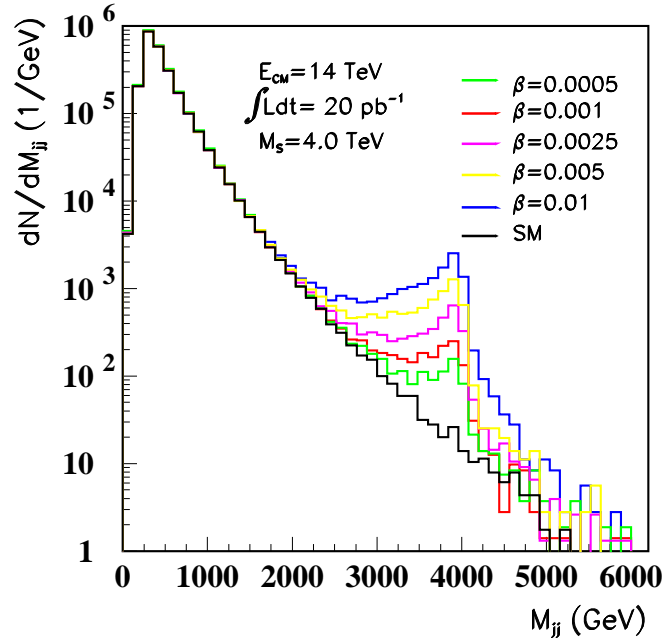}}
{\vskip -1.5cm}
\caption[]{\large {The influence of the first Regge recurrences on the  
dijet invariant mass distributions in $pp$ interactions for different
$\beta$ (see the text).
$E_{CM}=14$ TeV, $M_S=4$ TeV.}}
\label{Norm}
\end{center}
\end{figure}

The upper  panels in  Fig.~1 present the                    
dijet invariant mass distribution 
predictions for the subprocesses with two quarks and two gluons
for different $M_S$ and
$E_{CM}=7$ and 14 TeV. The lower panels show
the same for the $gg \to gg$ subprocess only for  $M_S=3$ and 4 TeV 
at  $E_{CM}=7$ and 14 TeV, respectively.
The  relation between
$\hat s $ and $M_S^2$ is  not regulated in any way.    
Comparing the plots in Fig.~1 we can see that string Regge 
recurrences in the $gg \to gg$ subprocess lead to unreasonable high
cross sections throughout the $M_{jj}$ interval if we do   not     
restrict  $\hat s$. Even for the 
subprocesses with two quarks and two gluons the resonance
contributions do not disappear towards to high $M_{jj}$ far 
from $M_S$.

Further we study the effect of restriction of  ${\hat s}$ on the
cross sections of string-induced quark-gluon and
four-gluon  events.
The blue(red) curve     in Fig.~2 is  the total
cross section of  $pp$ interactions for  
(6)-(8)((9))  channels as a function
of ${\alpha}^2$, where $ {\hat s} \ge {\alpha}^2 M_S^2$. 
The lower bounds on $\hat s$
range  from $\hat s_{min}$ to $M_S^2$.
The left(right)
panel in Fig.~2 is for $E_{CM}=7(14)$ TeV. 
There is no ${\alpha}^2$ dependence
for the subprocesses with two quarks and two gluons. The cross section
of the $gg \to gg$ channel  vary by more than two 
orders of magnitude with ${\alpha}^2$.

To avoid confusion we will limit ourselves to the region
near the first Regge recurrence  threshold 
$\hat s \approx M_S^2$.
In Fig.~3 we plot the dijet mass distributions for different       
upper bounds of the  variable
\begin{equation}
 \beta  = { {|(\hat s - M_S^{2})|}\over {\hat s}}. 
\end{equation}
Figure~3 displays the contributions of subprocesses (4)-(9) 
to the $pp$ interaction 
for $E_{CM}$=$14$ TeV and $M_S$=$4$ TeV. The SM prediction is also shown.
The resonances are easily observable above the SM background in the
wide range of $\beta$. As expected, the signal increases with $\beta$.
%%%%%%%%%%%%%%%%%%%%%%%%%%%5
%%\begin{figure}[htbp]
\begin{figure}[h!]                             %%   [H]      
\begin{center}
{\vskip -0.0cm}
{\hskip  0.0cm}  {\epsfxsize  5.0 truein \epsfbox{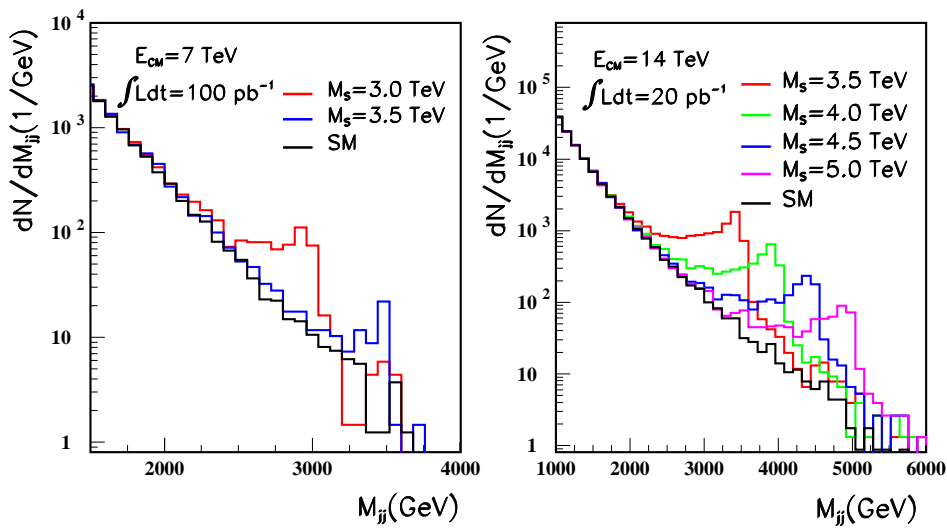}}
{\vskip -4.5cm}
\caption[]{\large {The expected  dijet invariant mass distributions for 
$pp$ interactions involving all subprocesses with the first Regge recurrences 
for different $M_S$ and                            
$\beta=0.0025$.       
The left(right) panel corresponds to  $E_{CM}=7(14)$~TeV.}}
\label{Norm}
\end{center}
\end{figure}

In Fig.~4 we show  the influence of string contributions         
from subprocesses (6)-(9) to the dijet invariant mass distribution 
for the upper bound  of $\beta=0.0025$ and different $M_S$. 
This  choice  of $\beta$ comes from the 
$\sim$$5\%$ resolution of the dijet mass  for ATLAS near 5 TeV.  
The left(right) panel in Fig.~4 corresponds to $E_{CM}=7(14)$~TeV.
The corresponding signal significances $S /\sqrt(S+B)$, where
the signal~(S) and background~(B) rates estimated in the invariant
dijet mass window  $[M_S-2\Gamma ,M_S+2\Gamma]$, are demonstrated  
in Table~1. The string resonance decay  widths 
into SM particles  $\Gamma$
were  calculated from the string amplitudes in\cite{anch3}.
We find that
the first  Regge recurrences at  $M_S=3.5(5)$ TeV might be 
observed   with  the  early data of 170(2) pb$^{-1}$ 
at  $E_{CM}=7(14)$ TeV      
with significance $>5\sigma$. 

\vspace*{0cm}
\begin{table}[]
{\normalsize
\begin{center}
{\color{black}        
\begin{tabular}{||c||c||c||c||c||c||}       \hline \hline
      
$E_{CM}=7\,TeV $&$       7\,TeV$ & $       14\,TeV$& $       14\,TeV$ & $       14\,TeV $& $       14\,TeV$  \\  
$M_S=3\,TeV  $&$    3.5\,TeV$ & $    3.5\,TeV$ & $    4\,TeV$  &$    4.5\,TeV $&$    5\,TeV$\\ \hline \hline
$S/{\sqrt{S+B}}=15.2    $&$4.0   $ & $83.2   $ &$50.1   $  &$31.5  $&$20.6$ \\ \hline \hline  
 
\end {tabular}
\caption[]{\large{Signal significances for the first Regge  recurrences for  the early LHC data
of 100(20) pb$^{-1}$ and $E_{CM}=7(14)$ TeV.}} 
\label{Norm}
}
\end{center}
}
\end{table}

%%%%%%%%%%%%%%%%%%%%%%%%%%%5
%%\begin{figure}[htbp]
\begin{figure}[h!]                             %%   [H]      
\begin{center}
{\vskip -2.0cm}
{\hskip  0.0cm}  {\epsfxsize 5.0 truein \epsfbox{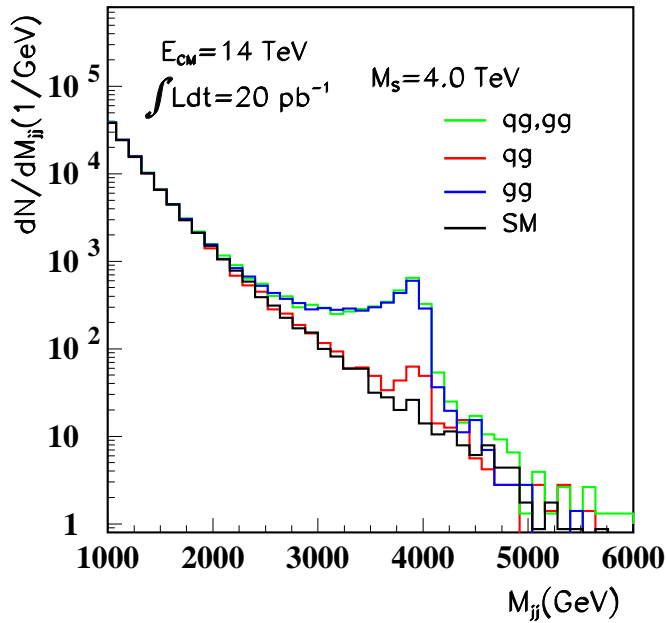}}
{\vskip -1.5cm}
\caption[]{\large { The dijet invariant mass distributions for $pp$ interactions
with all string-influenced subprocesses(green),  quark-qluon(red),   
and four-gluon(blue) subprocesses. $E_{CM}=14$~TeV, $M_S=4$~TeV.}}
\label{Norm}
\end{center}
\end{figure}

The ratios of the signal events corresponding to different
values of $\beta$ (see Fig.~3) are 
$$ 0.2:0.4:1.0:2.1:4.1.$$ 

In Fig.~5 we plot separate   and simultaneous influence of
low mass string production  of the                   
quark-gluon and four-gluon subprocesses   on the dijet
invariant mass distribution in $pp$ interactions at 14~TeV 
and $M_S=4$~TeV.
The SM predictions are also 
demonstrated in Fig.~5.
It is obvious that the four-gluon  signal strongly dominates the
quark-gluon signal.
The respective    signal significances for subprocesses (4)-(8), (9),
(4)-(9) are 5.14, 48.8 and 50.1.

\section{Conclusions}                                            
$ $

\vskip -0.5cm
High energy experimental physics  again becomes a wonderful
arena to look at the past, present  and future of our universe.
The ATLAS and CMS detectors now observe  particle collisions
at the energies never reached  before.

The string theory could be  the ultimate description of Nature,
but it is not  experimentally substantiated yet. 
In this paper, we have examined the dijet production at LHC 
as a window  to  the discovery   of low mass strings.

We show that if the string scale is low, the model-independent
part of string signatures, namely, processes involving four 
gauge bosons or two bosons and two fermions,    
could be manifested through resonances in the dijet invariant  mass
distributions in the early LHC data.
The major  part of the resonances comes from the
subprocesses with four bosons.
We find that  the LHC                       
is capable of probing  the string mass scale $\ge$3.5(5) TeV
with only 170(2) pb$^{-1}$ of integrated luminosity 
at $E_{CM}=7(14)$ TeV.

The low mass string scenario correctly explains\cite{anch4}
the anomaly  in the dijet invariant mass distributions of 
dijet events produced in association with a $W$ boson\cite{tevwjj}.

Another signature of the low string scale and large extra 
dimensions would be the discovery of mini black holes.
}
\

}}
\end{document}